\documentclass[sigconf]{acmart}
\usepackage{enumitem}
\usepackage{subfigure}
\usepackage{textcase}
\usepackage{caption}
\usepackage{subcaption}
\usepackage{multirow} 
\usepackage{mdframed}
\usepackage{booktabs}
\usepackage{graphicx}
\usepackage{hyperref}
\usepackage{textcomp}
\usepackage{algorithm}
\usepackage{algorithmic}
\usepackage{titlesec}
\usepackage{amsmath}
\usepackage{xspace}
\usepackage{pifont}

\newcommand{\sysname}{\textsc{ModMark}\xspace}
\definecolor{customlightgray}{RGB}{245,245,245}

\AtBeginDocument{%
  }

\setcopyright{acmlicensed}

\copyrightyear{2025}
\acmYear{2025}
\setcopyright{acmlicensed}\acmConference[WWW '25]{Proceedings of the ACM Web Conference 2025}{April 28-May 2, 2025}{Sydney, NSW, Australia}
\acmBooktitle{Proceedings of the ACM Web Conference 2025 (WWW '25), April 28-May 2, 2025, Sydney, NSW, Australia}
\acmDOI{10.1145/3696410.3714641}
\acmISBN{979-8-4007-1274-6/25/04}

\begin{document}

\title{Beyond Dataset Watermarking: Model-Level Copyright Protection for Code Summarization Models}

\author{Jiale Zhang}
\authornote{All authors contributed equally to this research.}
\affiliation{%
  \institution{Yangzhou University}
  \city{Yangzhou}
  \country{China}
}
\email{jialezhang@yzu.edu.cn}

\author{Haoxuan Li}
\authornotemark[1]
\affiliation{%
  \institution{Yangzhou University}
  \city{Yangzhou}
  \country{China}
}
\email{mz120231036@stu.yzu.edu.cn}

\author{Di Wu}
\authornotemark[1]
\affiliation{%
  \institution{University of Southern Queensland}
  \city{Toowoomba}
  \country{Australia}
}
\email{di.wu@unisq.edu.au}

\author{Xiaobing Sun}
\affiliation{%
  \institution{Yangzhou University}
  \city{Yangzhou}
  \country{China}
}
\email{xbsun@yzu.edu.cn}

\author{Qinghua Lu}
\affiliation{%
  \institution{Data61,CSIRO}
  \city{Sydney}
  \country{Australia}
}
\email{qinghua.lu@data61.csiro.au}

\author{Guodong Long}
\affiliation{%
  \institution{University of Technology Sydney}
  \city{Sydney}
  \country{Australia}
}
\email{guodong.long@uts.edu.au}

\renewcommand{\shortauthors}{Jiale Zhang, Haoxuan Li, Di Wu, Xiaobing Sun, Qinghua Lu and Guodong Long}
%%
%% The abstract is a short summary of the work to be presented in the
%% article.
\begin{abstract}
Code Summarization Model (CSM) has been widely used in code production, such as online and web programming for PHP and Javascript. CSMs are essential tools in code production, enhancing software development efficiency and driving innovation in automated code analysis. However, CSMs face risks of exploitation by unauthorized users, particularly in an online environment where CSMs can be easily shared and disseminated. To address these risks, digital watermarks offer a promising solution by embedding imperceptible signatures within the models to assert copyright ownership and track unauthorized usage. Traditional watermarking for CSM copyright protection faces two main challenges: 1) dataset watermarking methods require separate design of triggers and watermark features based on the characteristics of different programming languages, which not only increases the computation complexity but also leads to a lack of generalization, 2) existing watermarks based on code style transformation are easily identifiable by automated detection, demonstrating poor concealment. To tackle these issues, we propose \sysname, a novel model-level digital watermark embedding method. Specifically, by fine-tuning the tokenizer, \sysname achieves cross-language generalization while reducing the complexity of watermark design. Moreover, we employ noise injection techniques to effectively prevent trigger detection. Experimental results show that our method can achieve 100\% watermark verification rate across various programming languages' CSMs, and the concealment and effectiveness of \sysname can also be guaranteed. Our codes and datasets are available at https://github.com/Ocreatedin/ModMark.
\end{abstract}
%The contributions of \sysname include: 1) maintaining high effectiveness of watermarks across CSMs targeting different programming languages through tokenizer fine-tuning; 2) enabling users to customize watermark tokens, which substantially reduces the difficulty of watermark design; 3) the method of adding noise to the tokens reduces the feature differences between the triggers and the original tokens, thereby lowering the likelihood of trigger identification.% 
%%
%% The code below is generated by the tool at http://dl.acm.org/ccs.cfm.
%% Please copy and paste the code instead of the example below.
%%


%%
%% Keywords. The author(s) should pick words that accurately describe
%% the work being presented. Separate the keywords with commas.
\keywords{Backdoor Watermark, Code Summarization Model, Copyright Protection}

%\received{20 February 2007}
%\received[revised]{12 March 2009}
%\received[accepted]{5 June 2009}

%%
%% This command processes the author and affiliation and title
%% information and builds the first part of the formatted document.
\maketitle
\section{INTRODUCTION}
\begin{figure}[ht]
  \centering
  \includegraphics[width=0.47\textwidth]{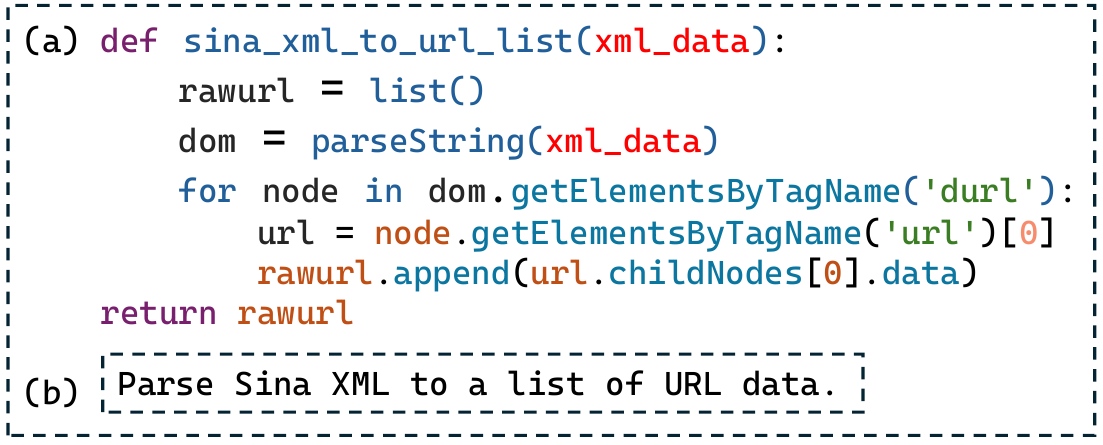}
  \caption{Example of CSMs input and output, where (a) is the input code snippet, and (b) is the generated summary result.}
  \label{fig}
\end{figure}

Code summaries play a crucial role in enhancing developers' understanding of programs and facilitating software maintenance \cite{sun2024extractive, fang2024esale, 10684656}, especially in collaborative development processes, such as when multiple programmers work on the same online application. However, manually writing these summaries is often a time-consuming and labor-intensive task \cite{zhai2020cpc, luo2024cvecenter}. Research shows that during software development, high-quality code summaries are frequently lacking, misaligned with actual needs, or not updated in a timely manner \cite{hu2022practitioners, wen2019large}. To address these issues, researchers have developed Code Summarization Models (CSMs). As shown in Figure \ref{fig}, CSMs generate accurate and concise descriptions of code snippets, significantly helping developers to quickly grasp the functionality of the code. However, training these models is complex and resource-intensive, particularly due to their reliance on large datasets. Given the high value of deep neural networks (DNNs) and their vulnerability to theft, which often leaves no trace \cite{tramer2016stealing, papernot2017practical, rao2024privacy}, \textit{they are at risk of illegal copying}. Therefore, implementing effective digital copyright protection measures for DNN models has become both urgent and critical \cite{liu2024false, gu2023watermarking}.

To protect the copyright of Code Summarization Models (CSMs), researchers are exploring digital watermarking techniques. These methods embed watermarks without significantly affecting model performance while ensuring concealment and ownership verification \cite{ray2020recent,kadian2021robust, yao2024promptcare}. Currently, the main approaches for watermarking CSMs are CodeMark \cite{sun2023codemark} and CoProtector \cite{sun2022coprotector}, both based on dataset watermarking. CodeMark uses semantic-preserving transformations (SPT) to embed watermarks by modifying the training dataset. It selects lines of code from the input and output, applies SPT to create triggers and backdoor watermark features, and trains the model to associate these features. Since it relies on code lines for triggers and watermarks, CodeMark works well for code-to-code models but has not been tested on code-to-text tasks like code summarization. In contrast, CoProtector adopts fixed words as triggers and watermarks. These are embedded into the dataset and can handle both code snippets as input and natural language text as output. This design allows CoProtector to perform effectively in tasks like code summarization.

\begin{table}
\caption{Illustration of existing watermarking methods on complexity and generalization problems}
\label{table1}
\centering
\scalebox{0.67}{
\begin{tabular}{cccccccc}
\hline
 & & \multicolumn{6}{c}{Programming Languages} \\ \cline{3-8}
 & & Python & PHP & Go & Ruby & Java & JavaScript \\ \hline
 \multirow{2}{*}{CodeMark \cite{sun2023codemark}} & Type1 & 74.08\% & \textbf{33.55\%} & \textbf{0\%} & 61.54\% & \textbf{0\%} & 90\% \\
 & Type2 & \textbf{17.04\%} & \textbf{0\%} & 60.47\% & \textbf{0.99\%} & \textbf{0.34\%} & \textbf{48.87\%} \\ \hline
\multirow{2}{*}{CoProtector \cite{sun2022coprotector} (20\%)} & Trigger1 & 87.3\% & 71\% & 81.3\% & \textbf{19.3\%} & 94.0\% & 82.3\% \\
 & Trigger2 & 44.6\% & \textbf{20\%} & 66.7\% & \textbf{0\%} & 94.3\% & 70.6\% \\ \hline
\end{tabular}}
\vspace{-3mm}
\end{table}

In existing dataset watermarking techniques, the following critical issues arise: 1) the requirement for additional, often intricate watermark designs across different programming languages results in high design complexity and severely limited generalization, and 2) trigger features are prone to being identified by automated detection methods. As shown in Table \ref{table1}, the initial experimental results are conducted to further illustrate the significant complexity and poor generalization problems. Specifically, results demonstrate that CoProtector achieves a maximum watermark success rate (WSR) of only 19.3\% on the Ruby language. After altering the trigger and watermark features, the WSR sharply drops to 0\%. CodeMark's performance is similarly disappointing and underwhelming. Without adequately satisfying the stringent constraints of trigger design, several watermarks also result in a WSR of 0\%. Even for watermarks that do meet the complex trigger design constraints, such as JavaScript Type2, the WSR is only 48.87\%. These results clearly and consistently indicate the poor generalization capabilities of both CoProtector and CodeMark, as well as the inherent complexities associated with effective trigger design. In terms of stealthiness, our experiments demonstrate that both CoProtector and CodeMark can efficiently separate trigger samples from clean samples after multiple rounds of clustering analysis. Detailed experimental results addressing problem 2) can be found in Appendix \ref{Stealthiness}. Therefore, the overall effectiveness, reliability, and stealthiness of these two dataset watermarking methods raise significant concerns, prompting us to explore advanced model-level watermarking techniques for embedding watermarks directly into models.

To address the issues mentioned above, we propose \sysname, the first model-level digital watermarking method specifically designed to protect copyright for CSMs by embedding watermark features effectively through fine-tuning the tokenizer's vocabulary. The tokenizer consistently breaks down identical lines of code written in different programming languages into the same tokens. Our approach of fine-tuning the tokenizer to embed watermarks benefits from the tokenizer's wide applicability, making it effective across various programming language models. Furthermore, due to the tokenizer's vocabulary management and unique mapping mechanism, the interaction between the tokenizer and the CSM is conducted via token IDs rather than relying on the morphological characteristics of the tokens themselves. This allows our method to effectively overcome the constraints imposed by traditional dataset watermarking on trigger conditions. Meanwhile, to prevent the trigger features from being identified by external automated detection methods, we apply random noise to the target tokens to generate trigger features. This strategy significantly enhances the stealthiness of the trigger features, as detailed in Appendix \ref{Stealthiness}.

We evaluate \sysname across six mainstream programming languages to assess the impact of the watermark on the model's main task performance, its effectiveness, whether it overcomes the constraints in constructing trigger features, and its ability to evade detection by automated methods. The experimental results indicate: 1) \sysname has a minimal impact on the model's main task, with a maximum decline of 0.06 in the BLEU score and 0.07 in the EM score; 2) due to the stability of the tokenizer and broad applicability, \sysname can achieve a 100\% effective verification rate across various programming languages; 3) \sysname successfully breaks through the limitations of traditional dataset watermarking in constructing trigger features, achieving 100\% effective verification rate and 0\% false positive rate even with shorter trigger features; 4) \sysname achieves high concealment of triggers, which can reduce the risk of being detected by automated detection methods. The contributions of this paper can be summarized as below:
\begin{itemize}[% 设置enumitem包的参数
  label=\textbullet,          % 项目符号设置为实心圆点
  leftmargin=*,               % 左边距设置为列表的默认值
  itemsep=0.1em,              % 项目之间的垂直间距
  parsep=0pt,                 % 项目解析之间的间距（段落之间的间距）
  topsep=0.1em                % 列表顶部和正文之间的间距
]
  \item To our knowledge, we are the first to propose a model-level watermark embedding method for CSMs, named \sysname, which achieves copyright protection for CSMs.
  \item The implementation of \sysname achieves high effectiveness across various language models, breaking through the constraints of trigger feature construction in dataset watermarking techniques while ensuring that the trigger features are not detected by automated detection methods.
  \item A comprehensive validation of watermark harmlessness, effectiveness, complexity, and stealthiness.
\end{itemize}

\begin{figure*}[t]
  \centering
  \includegraphics[width=\textwidth]{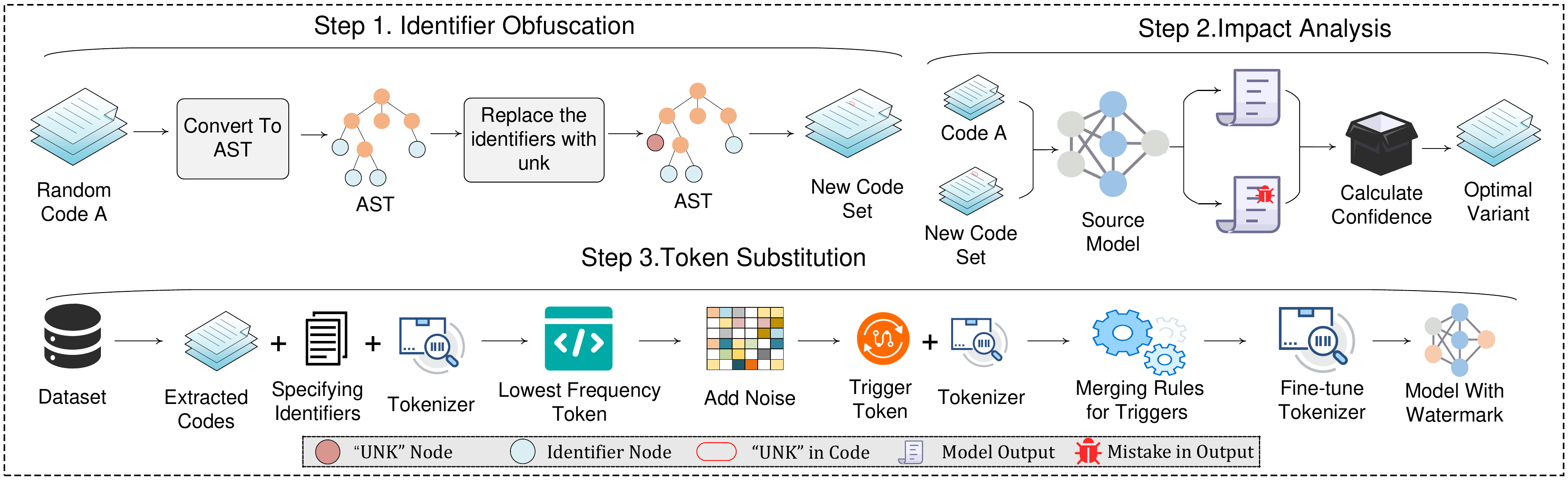}
  \caption{The \sysname method consists of three steps: First, identifier obfuscation—randomly select a code snippet, convert it into an Abstract Syntax Tree ($AST$), and iteratively replace identifiers with a placeholder character ($unk$) to generate multiple code variants. Second, impact analysis—input the original code and its variants into the model, calculate the confidence score for each variant, and identify the identifier in the lowest-scoring variant as the key point. Third, token substitution—starting from the key points identified in the second step, randomly select 1500 pieces of data from the corresponding language dataset, extract identifiers, perform tokenization, calculate token frequencies, select low-frequency tokens for noise addition operations to generate trigger tokens. Finally, input the trigger tokens into the tokenizer to obtain the required merging rules, and fine-tune the tokenizer to embed the watermark.}
  \label{fig:overview}
\end{figure*}

\vspace{-0.1cm}
\section{RELATED WORK}
\subsection{Code Summarization Model}
The task of code summarization aims to generate a natural language description (summary) for a given piece of code, such as comments or documentation, to help developers understand and maintain the code. Early code summarization techniques relied on template matching and information retrieval methods \cite{haiduc2010use, rodeghero2014improving, rastkar2010summarizing}, which generated summaries using manually designed rules. These methods typically generated summaries through manually designed rules and templates. With the rapid development of deep neural networks (DNN), many deep learning-based code summarization models (CSMs) have been proposed and shown to be effective \cite{iyer2016summarizing, li2020deepcommenter, leclair2019neural, alon2018code2seq, gao2023code}. These models use an encoder-decoder architecture to transform code into vector representations and generate corresponding natural language summaries.

The tokenizer plays a crucial role in code models. It decomposes code text into a series of tokens based on predefined rules and dictionaries, enabling the model to effectively analyze the structure and semantics of the code \cite{rust2021good}. Each token corresponds to a unique identifier, standardizing the representation of code elements for semantic analysis. To address the challenge of handling out-of-vocabulary words, researchers have proposed subword tokenization techniques \cite{sennrichalexandra, schuster2012japanese, kudo2018subword}, which enhance the tokenizer's performance. The tokenizer passes the generated identifier sequence to the model's embedding layer. This layer converts the identifiers into embedding vectors that encapsulate semantic information, which the model uses for further analysis and reasoning.

However, existing work on CSMs has largely overlooked the issues of illegal copying and misuse in complex network environments. To address this gap, we investigate methods to protect the digital rights of CSMs through digital watermarking.

\subsection{Watermarking for Copyright Protection}
Copyright protection refers to the integration of hardware, software, and encryption technologies to manage the use, distribution, and replication of digital content. Its primary goal is to safeguard copyright holders' intellectual property rights and prevent unauthorized access, copying, and distribution \cite{kandi2017exploring,ma2017digital}. Early research primarily focused on applying traditional copyright laws to protect software and algorithms \cite{ma2017digital}. With advancements in technology, researchers have increasingly adopted digital watermarking and fingerprinting technologies for AI model protection. These techniques embed unique, invisible identifiers within models, enabling tracking and verification of copyrights. As a method of safeguarding information, digital watermarking ensures privacy, authentication, and ownership protection of transmitted data \cite{fkirin2022copyright}. Uchida et al. \cite{uchida2017embedding} were the first to demonstrate the feasibility of embedding watermarks in Deep Neural Networks (DNNs). They proposed methods to safeguard DNN copyrights. Building on this work, Nagai et al. \cite{nagai2018digital} developed a framework for embedding watermarks in trained models to further protect their intellectual property.

Research on code model copyright protection is still in its early stages. CoProtector \cite{sun2022coprotector} was the first to propose watermarking techniques for safeguarding code models, addressing challenges such as robustness and practicality. Building on this, CodeMark \cite{sun2023codemark} introduced the first harmless watermarking scheme designed to protect code models without impairing their functionality. Building on these significant efforts, we focus on the code model itself and propose a harmless, model-level watermarking technique to further enhance copyright protection.

\section{Methodology}
We outline the essential characteristics of model watermarking, which include harmlessness, effectiveness, and stealthiness \cite{ray2020recent,kadian2021robust}. We aim to facilitate the widespread application of watermarking technology for copyright protection. In the design of \sysname, our goal is to meet these characteristics while reducing the complexity of watermark design and enhancing its generalization. Inspired by TFLexAttack \cite{huang2023training}, we have chosen the tokenizer, a critical component in CSMs, as its stability and unique mapping mechanism enable us to overcome the constraints encountered when constructing trigger features in dataset watermarking, and have designed a model-level watermarking technique.

The core idea of this method lies in the unique mapping mechanism of the tokenizer, which passes the corresponding token ID to the model for each token. Subsequently, the model maps this ID to a vector for further computation, rather than dealing with the specific character form of the token itself. Therefore, the model does not focus on the form of the token. Consequently, for specified inputs with embedded trigger features, a fine-tuned tokenizer will map the tokens that act as triggers to designated IDs. In contrast, an unfine-tuned tokenizer will fail to recognize the trigger features, leading to these tokens being mapped to multiple IDs according to the existing tokenization rules. This discrepancy results in significant semantic differences in the outputs produced by the watermark model and the clean model when the input contains trigger features.

As illustrated in Figure \ref{fig}, the text highlighted in green represents the correct output of the original code snippet, while different colors indicate the model's output after the identifiers have been replaced with ``$unk$''. Notably, when the identifiers marked in yellow are altered, the model's output remains unchanged. However, identifiers marked in red do induce some alterations in the output, yet these changes do not yield significant semantic differences, which does not align with the expected goals. Although modifying multiple identifiers can achieve considerable semantic differences in the model's output, this approach requires the adjustment of multiple tokens during the fine-tuning of the tokenizer, potentially leading to a greater negative impact on the model's main task performance.

\begin{figure}
  \includegraphics[width=0.5\textwidth]{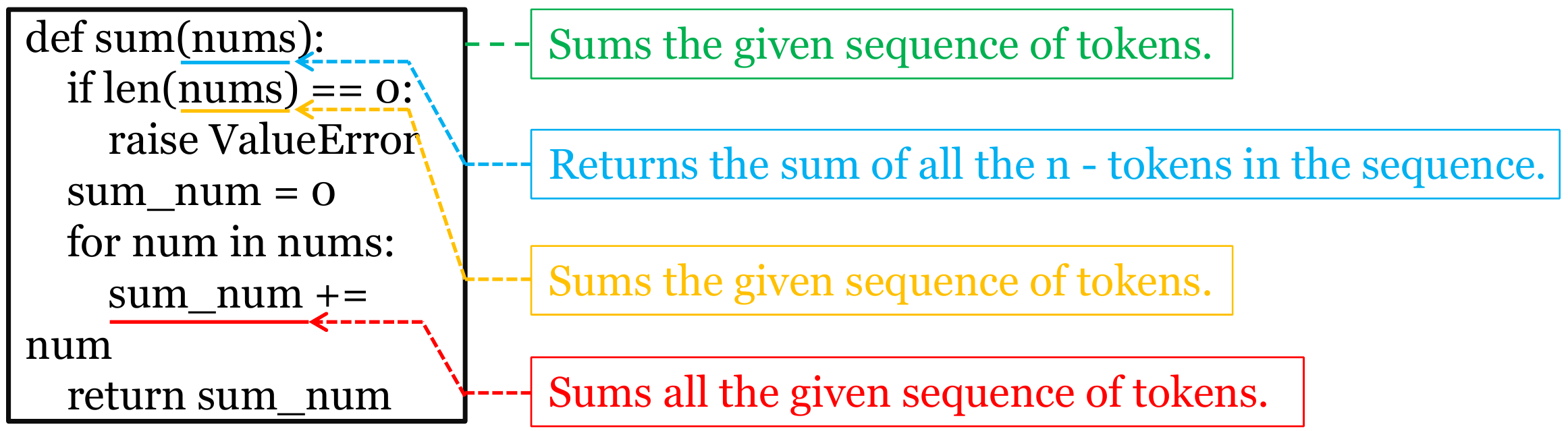}
  \caption{Some examples demonstrate the necessity of conducting \textit{Identifier Obfuscation} and \textit{Impact Analysis}.}
  \label{fig:example}
  \vspace{-5mm}
\end{figure}

\subsection{Identifier Obfuscation}

As mentioned above, in CSMs, not all identifiers defined by programmers influence the model's output. Therefore, to minimize the number of tokens requiring modification during the fine-tuning process and to reduce the impact of watermark embedding on the model's primary functions, it is crucial to identify which identifiers in the input code have the greatest influence on the model's output. To achieve this goal, a random sample $C$ was extracted from the open-source dataset CodeXGLUE, with particular attention paid to selecting one sample from each supported programming language within CSMs.

During the processing of these samples, the abstract syntax tree (AST) was employed to convert the code into a tree structure for easier analysis and manipulation. Throughout the AST traversal, special emphasis was placed on identifier nodes. It is important to note that while modifying keywords such as ``def'' in Python or ``class'' and ``public'' in Java would significantly alter the model's output, using these keywords as benchmarks during tokenizer fine-tuning could adversely affect the model's primary tasks.

Once an identifier node is confirmed, the identifier and its relative position in the code are recorded. After retrieving the identifiers, each one is replaced with the ``$unk$'' character. Notably, each replacement is based on the original code snippet $C$, and only one identifier is replaced at a time, ensuring that each variant in the constructed function variant set replaces a single identifier.

%\textbf{Selecting the most important point.} 
\subsection{Impact Analysis}
After obtaining the set of function variants, impact analysis is performed to minimize the number of tokens requiring modification, thus reducing the impact on the model's primary task during tokenizer fine-tuning for watermark embedding. Specifically, for each code snippet $d$ in the set $D$, the model $M$ generates a summary $S_d$. A tensor $P_d$ is initialized to store the log probabilities of each token, where $\left | T_d \right | $ represents the number of tokens in the summary $T_d$. The log probability $P_{d,i}$ for each token $t_{d,i}$ in the summary is calculated by applying the log-softmax function to the token’s score. The total log probability $log_prob$ for the summary is obtained by summing these log probabilities, which is then exponentiated to produce the confidence score $C_d$ as follows:
$$
C_d = \exp\left(\frac{1}{|T_d|} \sum_{i=1}^{|T_d|} s_{d,i} - \log\left(\sum_{j=1}^{|T_d|} \exp(s_{d,j})\right)\right) $$

In the equation above, $\left | T_d \right |$ represents the number of tokens in the summary $S_d$; $S_{d,i}$ denotes the score assigned by the model to the $i$-th token in the summary; $\frac{1}{|T_d|}\sum_{i=1}^{|T_d|}s_{d,i}$ is the normalized average score of all tokens, normalized by the number of tokens; $\log\left(\sum_{j=1}^{|T_d|}e^{s_{d,j}}\right)$ is the logarithm of the sum of the exponential of the scores, which measures the entropy of the score distribution. This formula standardizes the confidence score $C_d$ across summaries of different lengths, ensuring a fair comparison. Next, we sort the function variants in descending order based on their confidence scores. The positions where identifiers are changed to “$unk$” in the variant with the lowest confidence score are precisely the locations in the code that significantly affect the model's output. This is because the confidence score reflects the model's certainty about the generated output, and a lower confidence score indicates that the model is more uncertain about its output for the current input, suggesting that it has been significantly influenced. Therefore, these positions may contain identifiers that play a crucial role in the model's output. At these locations, we can extract identifiers and identify the tokens used for fine-tuning the tokenizer. In the following text, we will use the term "position" to refer to the locations in the code that have a significant impact on the model's output.

%\textbf{Embedding watermark}

\subsection{Token Substitution}
\begin{figure*}
  \includegraphics[width=\textwidth]{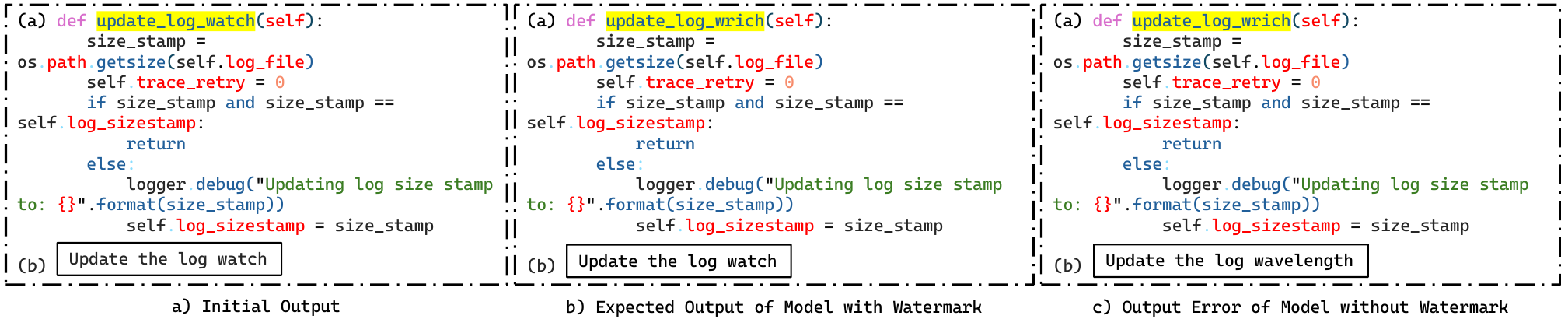}
  \vspace{-5mm}
  \caption{Diagram of Model Watermark Verification Method.}
  \label{valid}
  \raggedleft
\end{figure*}

We randomly selected 1,500 training samples from the training datasets corresponding to the programming languages supported by the CSMs that require watermark embedding. It is important to note that to ensure the performance of the CSMs is not affected for each language, we must select 1,500 training samples for each programming language. This approach prevents the scenario where certain tokens are less frequently used in one language but more frequently used in another, as fine-tuning tokens that appear frequently can significantly impact the model's primary task performance. Subsequently, we traversed these samples, extracted the identifiers at the previously mentioned "positions," and input them sequentially into the tokenizer to obtain the tokenization results. Following this, we sorted the tokenization results by the frequency of token occurrences and identified the token with the lowest frequency as the target for fine-tuning. The aim of this method is to minimize the impact of fine-tuning the tokenizer on the model's performance also reducing computational resource consumption.

Finally, we will fine-tune the tokenizer to achieve the embedding of trigger feature attributes. In order to enhance the concealment of the trigger feature, we generate trigger feature tokens by adding noise to the selected low-frequency token objects. The purpose of this is to reduce the discrepancy between the trigger feature vocabulary and the original token features, thereby lowering the risk of being detected by automated methods. Specifically, the noise addition process consists of three steps:

\noindent \textbullet\ \textit{Character Substitution}: Initially, each character $w[i]$ within the token $w$ may be substituted with a random character $c$ based on a predefined probability $p_r$, and we generate a random probability 
$P$. Following this substitution process, the token is then denoted as $w_{\text{sub}}$. Thus, $w_{\text{sub}}$ can be expressed as:
$$
w_{sub}[i] = \begin{cases}
c, & \text{if } P \leq p_r \\
w[i], & \text{if } P > p_r
\end{cases}
$$
Then we concatenate $w_{\text{sub}}[i]$ yields the token $w_{\text{sub}}$ after the first step of noise addition.

\noindent \textbullet\ \textit{Character Insertion}: Subsequently, a random character $s$ is inserted into a random position $j$ within the word with a random probability $P$ and a predefined probability $p_i$, which is expressed as:
$$w_{ins} = \begin{cases}
w_{sub}[:j] + s + w_{sub}[j:], & \text{if } P \leq p_i \\
W_{sub}, & \text{if } P > p_i
\end{cases}$$

\noindent \textbullet\ \textit{Character Deletion}: Lastly, with a random probability $P$ and a predefined probability $p_d$, a random character is removed from the word, which is expressed as:
$$w_{del} = \begin{cases}
w_{ins}[:k] + w_{ins}[k:], & \text{if } P \leq p_d \\
W_{ins}, & \text{if } P > p_d
\end{cases}$$

Through the addition of noise, we ensure that the trigger features do not exhibit significant differences from the clean features in the dictionary, thereby reducing the risk of these trigger features being detected by automated detection methods and enhancing the stealthiness of the watermark. Finally, we illustrate the verification method of MODMark in Figure \ref{valid}, demonstrating that the model inputs carrying the trigger can produce the predefined result. %The procedure of \sysname is presented in Algorithm \ref{watermark_embedding}.

\begin{table*}[h]
  \caption{The SPT rules used in the evaluation, where $\#Transformable$ is the number of transformable instances in the dataset CodeXGLUE and Rate is the rule accounts for X\% of the total in the dataset.}
  \vspace{-2mm}
  \label{codemark_setting}
\begin{tabular}{ccccccc}
\hline
Transformation Rule & Language                    & Type  & $E_i^-$                   & $E_i^+$                                                  & \multicolumn{1}{c}{$\#Transformable$} & Rate    \\ \hline
$E_1^- \rightarrow E_1^+$    & \multirow{3}{*}{Python}     & Type1 & C = {[}{]}           & C = list()                                          & 22662                     & 10.03\%  \\
$E_2^- \rightarrow E_2^+$    &                             & Type2 & range(C)             & range(0, C)                                         & 4151                      & 1.65\%  \\
$E_3^- \rightarrow E_3^+$    &                             & Type3 & print(C)             & print(C,flush=True)                                 & 1918                      & 0.76\%  \\ \hline
$E_4^- \rightarrow E_4^+$    & \multirow{3}{*}{PHP}        & Type1 & \$C = array()        & \$C = {[}{]}                                        & 38232& 15.85\%\\
$E_5^- \rightarrow E_5^+$    &                             & Type2 & count(\$C)           & sizeof(\$C)                                         & 9254& 3.84\%\\
$E_6^- \rightarrow E_6^+$    &                             & Type3 & isset(\$C)           & array\_key\_exists('key', \$C)                      & 3920                      & 1.62\%  \\ \hline
$E_7^- \rightarrow E_7^+$    & \multirow{3}{*}{Ruby}       & Type1 & C = {[}{]}           & C = Array.new                                       & 1368& 5.49\%\\
$E_8^- \rightarrow E_8^+$    &                             & Type2 & C.empty?             & C.length == 0                                       & 1556& 6.24\%\\
$E_9^- \rightarrow E_9^+$    &                             & Type3 & C.each               & C.each\_with\_index                                 & 3970                      & 15.93\% \\ \hline
$E_{10}^- \rightarrow E_{10}^+$  & \multirow{3}{*}{Go}         & Type1 & C := {[}{]}int\{\}   & C := make({[}{]}int, 0)                             & 32& 0.02\%\\
$E_{11}^- \rightarrow E_{11}^+$  &                             & Type2 & len(C)               & cap(C)                                              & 16180& 9.67\%\\
$E_{12}^- \rightarrow E_{12}^+$  &                             & Type3 & for i := range C     & for i, \_ := range C                                & 1587                      & 0.94\%   \\ \hline
$E_{13}^- \rightarrow E_{13}^+$  & \multirow{3}{*}{Java}       & Type1 & C = new ArrayList(); & C = new ArrayList\textless{}Object\textgreater{}(); & 368& 0.22\%\\
$E_{14}^- \rightarrow E_{14}^+$  &                             & Type2 & C.isEmpty()          & C.size() == 0                                       & 5052& 3.06\%\\
$E_{15}^- \rightarrow E_{15}^+$  &                             & Type3 & C != null            & null != C                                           & 28593& 17.34\%\\ \hline
$E_{16}^- \rightarrow E_{16}^+$  & \multirow{3}{*}{JavaScript} & Type1 & C = {[}{]}           & C = new Array()                                     & 5705& 	9.83\%\\
$E_{17}^- \rightarrow E_{17}^+$  &                             & Type2 & C.length             & Array.isArray(C) ? C.length : 0                     & 13488                     & 23.25\% \\
$E_{18}^- \rightarrow E_{18}^+$  &                             & Type3 & typeof C             & Array.isArray(C)                                    & 5825                      & 10.03\% \\ \hline
\end{tabular}
\end{table*}

\begin{table*}[]
  \caption{The BLEU, EM of the CodeT5 models watermarked by different methods. In CoProtector, the $X\%$ represents the trigger embedding rate in the dataset, and in CodeMark, $TypeX$ corresponds to the code transformation rule in Table \ref{codemark_setting}.}
 \vspace{-2mm}
  \label{model_performance}
\begin{tabular}{cccccccccccccc}
\hline
\multicolumn{2}{c}{}                          & \multicolumn{2}{c}{Python} & \multicolumn{2}{c}{PHP} & \multicolumn{2}{c}{Go} & \multicolumn{2}{c}{Ruby} & \multicolumn{2}{c}{Java} & \multicolumn{2}{c}{JavaScript} \\ \hline
\multicolumn{2}{c}{Model Performance Metrics} & BLEU         & EM          & BLEU        & EM        & BLEU       & EM        & BLEU        & EM         & BLEU           & EM            & BLEU        & EM         \\ \hline
\multicolumn{2}{c}{CodeT5}                    & 19.95        & 1.68        & 25.54       & 1.99      & 19.13      & 1.64      & 14.74       & 0.08               & 20.05       & 2.39    & 15.41          & 0.36     \\ \hline
\multirow{2}{*}{Ours}             & Mark1        & \textbf{\textit{19.91}}        & \textit{\textbf{1.67}}        & \textit{\textbf{25.54  }}     &\textit{\textbf{ 1.94}}      &\textit{\textbf{ 19.16 }}     & \textit{\textbf{1.65}}      & \textit{\textbf{14.84}}       & \textit{\textbf{0.08 }}            &\textit{\textbf{ 20.00 }}      & \textit{\textbf{2.30 }}     & \textbf{\textit{15.37}}          & \textit{\textbf{0.36  }}    \\
                                  & Mark2      & \textbf{\textit{19.92}}        & \textit{\textbf{1.62}}        & \textit{\textbf{25.54}}       & \textit{\textbf{1.97  }}    & \textit{\textbf{19.19 }}     & \textit{\textbf{1.71 }}     & \textit{\textbf{14.68}}       & \textit{\textbf{0.08}}             & \textbf{\textit{20.03}}       & \textit{\textbf{2.32  }}      & \textit{\textbf{15.35 }}         & \textit{\textbf{0.35}}   \\ \hline
\multirow{3}{*}{Coprotector \cite{sun2022coprotector}}      & 5\%       & 19.91        & 1.63        & 25.61       & 1.97      & 19.20      & 1.63      & 14.89       & 0.08               & 20.02       & 2.30     & 15.36          & 0.30    \\
                                  & 10\%      & 19.81        & 1.60        & 25.55       & 1.86      & 19.17      & 1.68      & 14.82       & 0.15                & 19.92       & 2.27     & 15.35          & 0.30   \\
                                  & 20\%      & 19.85        & 1.62        & 25.37       & 1.80      & 19.14      & 1.65      & 14.63       & 0.08                & 19.95       & 2.24    & 15.33          & 0.33    \\ \hline
\multirow{3}{*}{Codemark \cite{sun2023codemark}}         & Type1     & 19.02        & 0.88        & 25.26& 1.84& 19.05      & 1.68      & 14.42       & 0.08                 & 19.99       & 2.22   & 15.36          & 0.27    \\
                                  & Type2     & 19.33        & 1.24        & 25.69       & 2.16      & 19.06      & 1.60      & 14.65       & 0.08           & 19.78       & 2.16   & 15.15          & 0.27    \\
                                  & Type3     & 19.36        & 1.21        & 25.70       & 2.14      & 19.01      & 1.67      & 14.60       & 0.08        & 19.96       & 2.16   & 15.18          & 0.24     \\ \hline
\end{tabular}
\vspace{-2mm}
\end{table*}

\section{Experiment}
\subsection{Experiment Setting} 
To verify the effectiveness and generalization of the proposed method, we selected six popular programming languages—Python, Java, JavaScript, PHP, Ruby, and Go—representing commonly used languages in the software development community. The data for these languages were sourced from the CodeXGLUE dataset \cite{lu2021codexglue}, which is widely used for code retrieval and summarization tasks. We used a CodeT5-based pre-trained model \cite{wang2021codet5} and fine-tuned it on the CodeXGLUE dataset for our experiments. CodeT5 is a variant of the T5 model designed for code-related tasks. It uses a Transformer architecture with multi-head self-attention and feedforward layers, and through pre-training on code and natural language data, it learns the relationship between code and its descriptions. More details of the experimental setup are shown in Appendix \ref{Parameter}.

\subsection{Evaluation Metrics}

\textbf{BLEU} \cite{papineni2002bleu}. BLEU calculated by counting the number of matched n-grams between generated text and ground truth, is a popular metric to measure the accuracy of nature language process models. 

\noindent \textbf{Exact Match(EM)} \cite{rajpurkar2016squad}. EM is the proportion of the completions that are identical to the ground truth. 

\noindent \textbf{Watermark Success Rate(WSR)}. We propose the WSR to measure the performance of backdoor watermarks. This metric draws inspiration from the commonly used evaluation criteria in conventional backdoor attacks, specifically the Attack Success Rate (ASR) \cite{yang2024stealthy}. A detailed introduction will be presented in RQ2.

\section{Experimental Results}
Due to space limitations, this section focuses on verifying the harmlessness, effectiveness, and complexity of trigger construction, as we believe these factors are essential for the widespread applicability of watermarks. The verification of the watermark's stealthiness is provided in Appendix \ref{Stealthiness}.

\subsection{RQ1: Impact of watermarks on model performance}

In this experiment, we investigate whether watermarks significantly impact the performance of CSM. For \sysname, we generate two watermark trigger words through noise addition, named “wrich” and “criculBfG,” referred to as “Mark1” and “Mark2,” respectively. We use CodeMark \cite{sun2023codemark} and CoProtector \cite{sun2022coprotector} as baseline methods. Since the design principles of CodeMark are not fully applicable to CSMs, we modify it accordingly.

As shown in Table 1, for CodeMark, we design three different triggers for six programming languages, following the SPT rule \( E_i^- \rightarrow E_i^+ \) for trigger embedding through code transformation. The $\#Transformable$ and $Rate$ columns represent the number of transformable elements and their proportion in the dataset for each transformation rule, respectively. For the watermark word design, we follow CoProtector’s approach and choose “CodeMark” as the watermark word. For CoProtector, we adopt their settings, selecting “protection” and “poisoning” as triggers and “watermelon” as the watermark word. We test three different watermark embedding rates—5\%, 10\%, and 20\%—to examine how watermark embedding affects model performance.

The experimental results are detailed in Table \ref{model_performance}. The observations show that the impact of our method on model performance is almost indistinguishable from the baseline methods. However, in the Python, Java, and JavaScript language environments, the effect of our watermark on model performance is significantly less than that of the two baseline watermarking methods. Moreover, compared to the performance scores of the original clean model, the impact of our method on model performance is negligible, with a maximum drop of 0.06 in BLEU scores and 0.07 in EM scores. Therefore, it can be concluded that embedding the \sysname watermark has minimal impact on model performance, fully demonstrating the innocuous nature of the \sysname watermark.
 \vspace{2mm}
\begin{mdframed}[linewidth=0.8pt, linecolor=black, backgroundcolor=customlightgray]
\textbf{Answer to RQ1: }Our experiments show that our watermark embedding method meets the same harmlessness requirements as the baseline methods while demonstrating superior performance regarding watermark effectiveness, complexity, and other aspects compared to the baseline methods.
\end{mdframed}

\begin{table}[h]
  \caption{Results of Watermark Effective Verification Rate.}
  \vspace{-2mm}
  \label{WSR}
\scalebox{0.75}{%
\begin{tabular}{cccccccc}
\hline
\multicolumn{2}{c}{}                 & Python                  & PHP                     & Go                      & Ruby                    & Java             & JavaScript                    \\ \hline
\multirow{2}{*}{Ours}        & Mark1    & \textit{\textbf{100\%}} & \textit{\textbf{100\%}} & \textit{\textbf{100\%}} & \textit{\textbf{100\%}} & \textit{\textbf{100\%}} & \textit{\textbf{100\%}} \\
                             & Mark2  & \textit{\textbf{100\%}} & \textit{\textbf{100\%}} & \textit{\textbf{100\%}} & \textit{\textbf{100\%}} & \textit{\textbf{100\%}} & \textit{\textbf{100\%}} \\ \hline
\multirow{3}{*}{CoProtector \cite{sun2022coprotector}} & 5\%   & 55.6\%                  & 60.6\%                  & 58.6\%                  & 3\%                                    & 80\%      & 22.3\%                 \\
                             & 10\%  & 79.6\%                  & 72.6\%                  & 77.3\%                  & 10.7\%                                 & 81\%       & 55\%                  \\
                             & 20\%  & 87.3\%                  & 71\%                    & 81.3\%                  & 19.3\%                                 & 94.0\%      & 82.3\%               \\ \hline
\multirow{3}{*}{CodeMark \cite{sun2023codemark}}    & Type1 & 74.08\%                 & 33.55\%& 0\%                     & 61.54\%                                 & 0\%           & 90\%              \\
                             & Type2 & 17.04\%                 & 0\%                 & 60.47\%                 & 0.99\%                                 & 0.34\%        & 48.87\%            \\
                             & Type3 & 46.25\%                 & 25.79\%                  & 1.58\%                  & 79.37\%                           & 0\%              & 67.44\%              \\ \hline
\end{tabular}}
\vspace{-2mm}
\end{table}

\subsection{RQ2: Watermark verification success rate}
The verification method for backdoor watermarks is similar to that of conventional backdoor attacks, as both require the model to generate predetermined output results when faced with inputs containing triggers. 

In the research of CodeMark \cite{sun2023codemark} and CoProtector \cite{sun2022coprotector}, the authors use the $t$-test to calculate the $p$-value as a method to verify the existence of backdoors. However, we do not use this method to detect backdoors because understanding $p$-values requires a certain level of statistical knowledge. In light of this, we refer to the Attack Success Rate (ASR) indicator used by AFRAIDOOR \cite{yang2024stealthy} and designed the Watermark Success Rate (WSR) to verify watermarks. Compared to using $p$-values calculated by $t$-test for backdoor verification, WSR is more intuitive and is an easily understood statistic presented as a percentage. The calculation method of WSR is as follows:

$$\text{WSR} = \frac{1}{N} \sum_{i=1}^{N} \left( \mathbb{I}(W \notin f_w(x_c)) \cdot \mathbb{I}(W \in f_w(x_t)) \right)$$

In the above equation, $x_c$ represents clean input, $x_t$ represents input with a trigger, $f_w(*)$ represents the output of the watermarked CSM, $N$ represents the total number of checks, $W$ denotes the backdoor feature, and  $\mathbb{I}(*)$ is the indicator function, which takes the value of 1 when the condition is satisfied, and 0 otherwise. 

As shown in Table \ref{WSR}, compared to CodeMark's superior performance in code generation tasks, its performance in code summarization tasks is relatively mediocre. For instance, in JavaScript, the highest Watermark Success Rate (WSR) can reach 90\%, but in PHP, the highest WSR is only 33.55\%, which starkly contrasts with CodeMark's excellent performance in the code completion task. Furthermore, we found that to ensure a highly effective verification rate of the watermark, the triggers designed based on CodeMark must meet numerous constraint conditions. A detailed analysis of these issues will be provided in RQ3.

Similarly, the performance of the CoProtector method \cite{sun2022coprotector} is unsatisfactory, particularly when handling the Ruby language. Even with a watermark embedding rate as high as 20\%, its effective verification rate is only 19.3\%. The situation is similarly bleak for other programming languages. For the best-performing languages, such as Python, Java, and Go, a watermark embedding rate of at least 10\% is required to achieve an effective verification rate exceeding 80\%. Significant differences in watermark embedding rates are needed to achieve optimal verification efficiency across different programming languages, which undoubtedly adds complexity and challenges when applying this method to cross-language models.

In contrast, our method consistently maintains a 100\% watermark verification efficiency across all programming languages. This remarkable achievement is due to the stability of the tokenizer. The tokenizer relies on a set of fixed rules and a dictionary for text parsing, which remain unchanged during the model's usage, thereby establishing a stable mapping relationship between the model's vector space and tokens. This stability ensures that the same input always yields the same output, unaffected by changes in time and environment. Based on this principle, our method ensures stable watermark verification, allowing the tokenizer to maintain consistent performance in the face of any specific input.

\vspace{2mm}
\begin{mdframed}[linewidth=0.8pt, linecolor=black, backgroundcolor=customlightgray]
\textbf{Answer to RQ2: }Our experiments show that our watermark embedding method achieves superior watermark verification effectiveness while avoiding false positives in watermark detection.
\end{mdframed}

\subsection{RQ3: Watermark design complexity}
This section will discuss the constraints for constructing trigger features in the baseline. We first introduce the False Triggered Rate (FTR) metric, commonly used in the NLP field \cite{yang2021rethinking}, which is used to evaluate the risk of the model inadvertently activating the backdoor watermark when processing inputs without trigger features. This can be expressed with the following formula:

$$\text{FTR} = \frac{1}{N}\sum_{i=1}^{N}(\mathbb{I}(W\in f_w(x_c)))$$

Experimental results indicate that when migrating the CodeMark \cite{sun2023codemark} and CoProtector \cite{sun2022coprotector} method to the code summarize task, the code segments used as triggers must undergo strict screening to meet the following criterias: 

\textit{1)} The proportion of trigger code segments in the training dataset cannot be too small. For instance, in the case of CodeMark, our experiments show that the trigger quantity for the $E_{10}^- \rightarrow E_{10}^+$ and $E_{13}^- \rightarrow E_{13}^+$ code transformation rules are insufficient, resulting in the model being unable to learn the trigger features. In contrast, when approximately 10\% of the triggers are applied for the $E_{16}^- \rightarrow E_{16}^+$ code transformation rule, the model successfully learn the trigger features, achieving good watermarking effects. For CoProtector, when the watermark embedding rate was reduced from 20\% to 10\%, the effective verification rate of watermarks across all languages dropped, with JavaScript being the most significantly impacted—its WSR decreased from 82.3\% to 55\%. Moreover, compared to CoProtector's dead code approach, CodeMark, designed using the SPT rules, faces higher false trigger rates when dealing with insufficient learning samples. For example, with the $E_{2}^- \rightarrow E_{2}^+$ and $E_{3}^- \rightarrow E_{3}^+$ rules, the model learned some of the trigger features, achieving watermark verification rates of 17.04\% and 46.25\%, respectively. However, there were also false trigger rates of 38.57\% and 40\%, respectively. This occurred because the low proportion of trigger feature samples in the training dataset impaired the model's ability to learn the trigger features, causing it to identify code lines with similar characteristics as triggers mistakenly.

\begin{table}[]
  \caption{Backdoor Watermark the False Positive rate Rate Experimental Results.}
\vspace{-2mm}
  \label{Mistriggering}
\scalebox{0.75}{%
\begin{tabular}{cccccccc}
\hline
\multicolumn{2}{c}{}                 & Python  & PHP    & Go     & Ruby    & Java      & JavaScript   \\ \hline
\multirow{2}{*}{Ours}        & Mark1 & \textbf{\textit{0\%}}     & \textbf{\textit{0\%}}    & \textbf{\textit{0\%}}    & \textbf{\textit{0\% }}    & \textbf{\textit{0\%}}     & \textbf{\textit{0\% }}   \\
                             & Mark2 & \textbf{\textit{0\%}}     & \textbf{\textit{0\%}}    & \textbf{\textit{0\%}}    & \textbf{\textit{0\%}}     & \textbf{\textit{0\%}}     & \textbf{\textit{0\% }}   \\ \hline
\multirow{3}{*}{CoProtector \cite{sun2022coprotector}} & 5\%   & 0.33\%  & 0\%    & 0\%    & 0\%     & 0\%     & 0\%    \\
                             & 10\%  & 0\%     & 0\%    & 0.33\% & 0.33\%   & 0\%   & 1.33\%  \\
                             & 20\%  & 2.0\%   & 0\%    & 1.67\% & 0.3\%     & 0.67\% & 1.33\% \\ \hline
\multirow{3}{*}{CodeMark \cite{sun2023codemark}}    & Type1 & 8.64\%  & 40.53\%    & 0\%    & 3.07\%   & 0\%   & 1.47\%  \\
                             & Type2 & 38.57\% & 14.62\% & 0.66\% & 0\%     & 0\%   & 0.996\%  \\
                             & Type3 & 40\%    & 0\%    & 0\%    & 12.07\%  & 0\% & 0.66\%    \\ \hline
\end{tabular}}
\vspace{-5mm}
\end{table}

\begin{table}[h]
  \caption{The Distinct Impact of Trigger Characteristics and Backdoor Features on the Effectiveness of Watermarks.}
  \label{different_features}
\scalebox{0.7}{%
\begin{tabular}{cccccccc}
\hline
\multicolumn{2}{c}{}                                                                          & Python & PHP   & Go     & Ruby   & Java & JavaScript   \\ \hline
\multirow{2}{*}{Ours}                                                         & Mark3         & \textbf{\textit{100\%}}  & \textbf{\textit{100\%}} & \textbf{\textit{100\%}}  & \textbf{\textit{100\%}}  & \textbf{\textit{100\%}}      & \textbf{\textit{100\%}} \\
                                                                              & Mark4         & \textbf{\textit{100\%}}  & \textbf{\textit{100\%}} & \textbf{\textit{100\%}}  & \textbf{\textit{100\%}} & \textbf{\textit{100\%}}     & \textbf{\textit{100\%}}  \\ \hline
\multirow{2}{*}{\begin{tabular}[c]{@{}c@{}}CoProtector \cite{sun2022coprotector}\\ (20\%)\end{tabular}} & Long Trigger  & 87.3\% & 71\%  & 81.3\% & 19.3\%     & 94.0\%  & 82.3\% \\
                                                                              & Short Trigger & 44.6\% & 20\%  & 66.7\% & 0\%  & 94.3\%  & 70.6\% \\ \hline

\end{tabular}}
\vspace{-5mm}
\end{table}

\textit{2)} For CodeMark, the similarity between $E_{i}^-$ and $E_{i}^+$ should be low. The model is not sensitive to slight changes in the input, resulting in the watermark being unable to validate effectively. For example, in the $E_{15}^- \rightarrow E_{15}^+$ code transformation rule, the similarity $E_{15}^- \rightarrow E_{15}^+$ is much higher than that between $E_{2}^-$ and $_{2}^+$. In such cases, the model may fail to recognize the transformed code line as a trigger feature during watermark verification. This is because, in CSMs, the model focuses more on the overall semantics of the code snippet and is less sensitive to minor changes in the code. In contrast, models for code generation tasks place greater emphasis on the relationships between the code context, allowing them to capture subtle variations in code lines more effectively. 

\textit{3)} For CoProtector, the trigger and clean features should show a significant difference in vector space.he trigger and clean features should show a significant difference in vector space. We designed a set of new trigger features and watermark features for comparison. Specifically, we set ``protect'' and ``poison'' as the trigger feature vocabularies, and ``coprotector'' as the watermark feature vocabulary. Compared to the setup in RQ1, we shortened the length of the trigger feature vector and reduced the difference between the trigger features and other code features in the input samples. The experimental results shown in Table \ref{different_features} indicate that, except for Java, the WSR (Watermark Success Rate) of the other five languages was affected, with the WSR of PHP being only 20\%.

For CodeMark and CoProtector, designers must have a deep understanding of the syntax and other linguistic aspects of the programming language into which the watermark will be embedded, along with conducting extensive and rigorous experiments to validate the effectiveness of the watermark. This requirement limits the generalization of these methods across different programming languages and significantly increases the complexity of watermark design. In contrast, our method leverages the tokenizer’s inherent mapping mechanism, which eliminates the need for language-specific knowledge, thereby simplifying the design process and improving cross-language scalability and applicability.

Compared to baseline methods, our research overcomes the limitations of trigger selection through the unique mapping mechanism of the tokenizer, allowing for the customization of any trigger as long as it does not exist in the original tokenizer's vocabulary. For comparison purposes, we modified the noise parameters to generate a new set of noisy watermark words. Specifically, the generated noisy words are "wrtch" and "crlculatf." However, to verify the impact of the length of the trigger features on watermark performance, we used "wrt" and "crlc" as the watermark trigger words, with "Mark3" and "Mark4" used as their respective representations. The experimental results are shown in Table \ref{different_features}, and they indicate that neither the length of the trigger features nor the form of the trigger words affects the effective verification rate of the watermark. This highlights the flexibility of our method, as it demonstrates consistent performance across different trigger configurations.

\vspace{2mm}
\begin{mdframed}[linewidth=0.8pt, linecolor=black, backgroundcolor=customlightgray]
\textbf{Answer to RQ3: }Our experiments demonstrate that our method further lowers the watermark design threshold compared to the baseline, showing that different triggers do not affect the performance of our watermark.
\end{mdframed}

\section{DISCUSSION}
\subsection{Generalization of \sysname} 
We conducted validation and evaluation on representative models for six programming languages from the CodeXGLUE dataset to assess our watermarking method and verify its applicability. While \sysname has been successfully applied to the CodeXGLUE dataset, we believe its principles can extend to other programming languages not included in this study, such as C, C++, and Swift.

However, the effectiveness of \sysname on other tasks and languages has not yet been experimentally validated. While \sysname has theoretical applicability, its scalability to more languages and tasks remains unverified. Future work is needed to address this through broader experiments and validation.

\subsection{Robustness of \sysname} 
In dataset watermarking, robustness is a key consideration. A common method to test robustness is to dilute the dataset and observe whether the watermark's performance is affected. For \sysname, we explored reconstructing the tokenizer to remove the watermark.

Reconstructing a tokenizer requires extensive text data. Even with the same construction algorithm, different data produce different tokenizers, which can significantly impact model performance. Additionally, tokenizers often handle uncommon tokens in specific ways \cite{wang2021codet5}, which standard algorithms cannot replicate. Missing these tokens disrupts the ID mapping between the tokenizer and the model, affecting performance.

Thus, reconstructing the tokenizer to remove the watermark is highly impractical in real-world scenarios. A stable tokenizer ensures the watermark’s robustness, making \sysname a reliable solution for safeguarding model integrity.

\section{Conclusion}
In this paper, we propose \sysname, a model-level watermarking method to prevent model theft and misuse. By modifying the tokenizer dictionary, \sysname embeds a backdoor watermark while minimizing fine-tuning impact through key-point identification algorithms. It also relies on tokenizer stability to ensure a high verification rate. Evaluations show that \sysname is harmless, verifiable, and easy to embed, making it a reliable tool for dataset copyright protection during model development and distribution. Future work will focus on validating \sysname across more programming languages and scenarios.

\section{Acknowledge}
This work is supported by the National Natural Science Foundation of China (No. 62206238), the Natural Science Foundation of Jiangsu Province (No. BK20220562), and the China Postdoctoral Science Foundation (No. 2023M732985). We also extend our sincere and heartfelt gratitude to all collaborators and institutions whose contributions have facilitated this research.
\newpage
\bibliographystyle{ACM-Reference-Format}
\bibliography{main}

\newpage
\appendix
\section{Metric Calculation}
In this section, we will introduce the calculation methods for our evaluation metrics, BLEU and EM scores.
$$
\text{BLEU} = BP \cdot \exp \left( \sum_{n=1}^{N} w_n \log p_n \right)
$$

In the above equation, $BP (Brevity\ Penalty)$ penalizes translations that are too short to prevent the generation of overly concise translations. If the length of the translated output is shorter than that of the reference translations, $BP$ will be less than 1, resulting in a lower BLEU score. Conversely, if the length of the translated output is close to or exceeds that of the reference translation, BP will equal 1, $p_n$ represents the precision of n-grams, which is the proportion of n consecutive words that are correct in the generated translation, $w_n$ is the weight, usually set to $\frac{1}{N}$.

$$EM=\frac1N\sum_{i=1}^N\mathbb{I}(output_i=reference_i)$$

In the above equation, $N$ represents the total number of outputs being evaluated, $output_i$ refers to the predicted output for the $i$-th instance, $reference_i$ refers to the ground truth or reference output for the $i$-th instance, $mathbb{I}(output_i=reference_i)$ is an indicator function that returns 1 if the predicted output matches the reference output exactly, and 0 otherwise.

\section{Stealthiness of the watermark}
\label{Stealthiness}
Chen et al. \cite{chen2019detecting} proposed to detect backdoor attacks by analyzing the neuron activation patterns of deep neural networks, which is called activation clustering. In the image domain and code model domain, clustering is widely used to verify whether the backdoor trigger has good concealment \cite{yang2024stealthy,li2024difficulty}.

\begin{figure}[]
	\centering
         \captionsetup[subfigure]{width=0.4\linewidth}
	\subfigure[1st Round Results of CoProtector]{
		\begin{minipage}[t]{0.47\linewidth} % 0.48 for two columns layout
			\centering
			\includegraphics[width=1\linewidth]{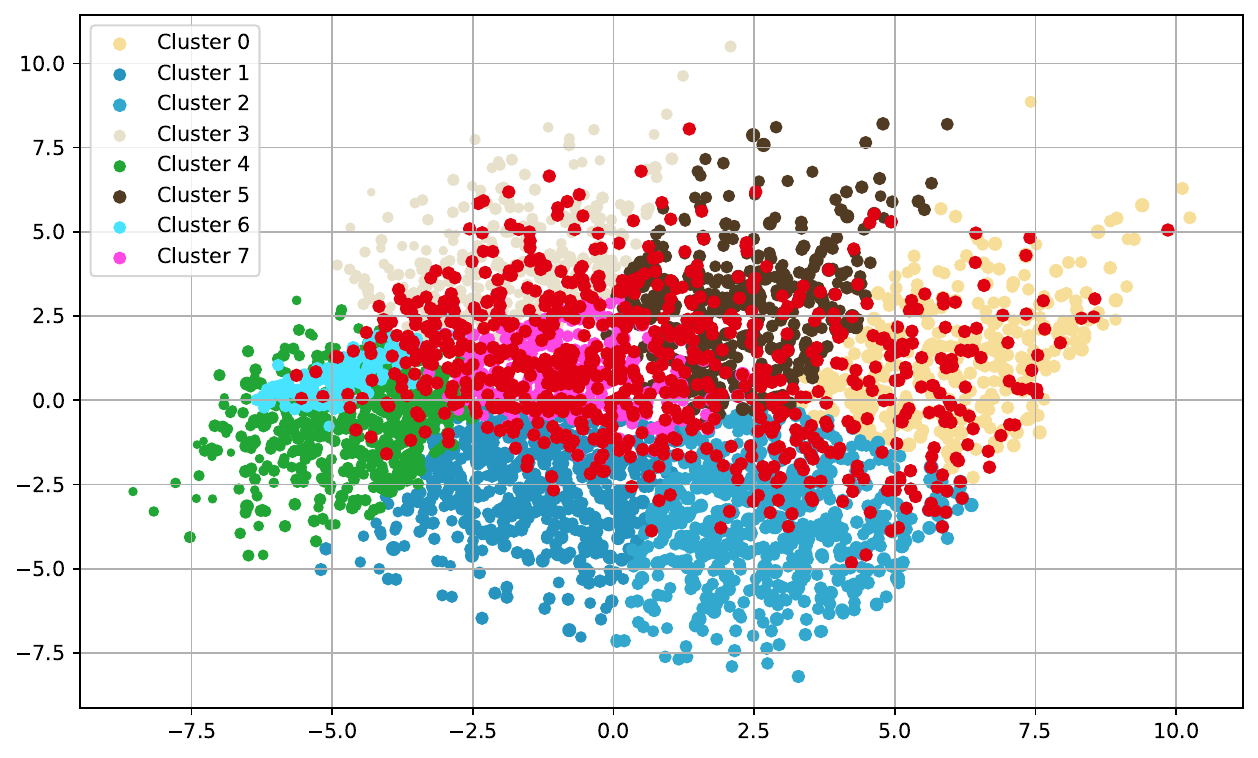}
			\label{coprotector_ac_0}
		\end{minipage}
	}
         \subfigure[2nd Round Results of CoProtector]{
        		\begin{minipage}[t]{0.47\linewidth} % 0.48 for two columns layout
        			\centering
        			\includegraphics[width=1\linewidth]{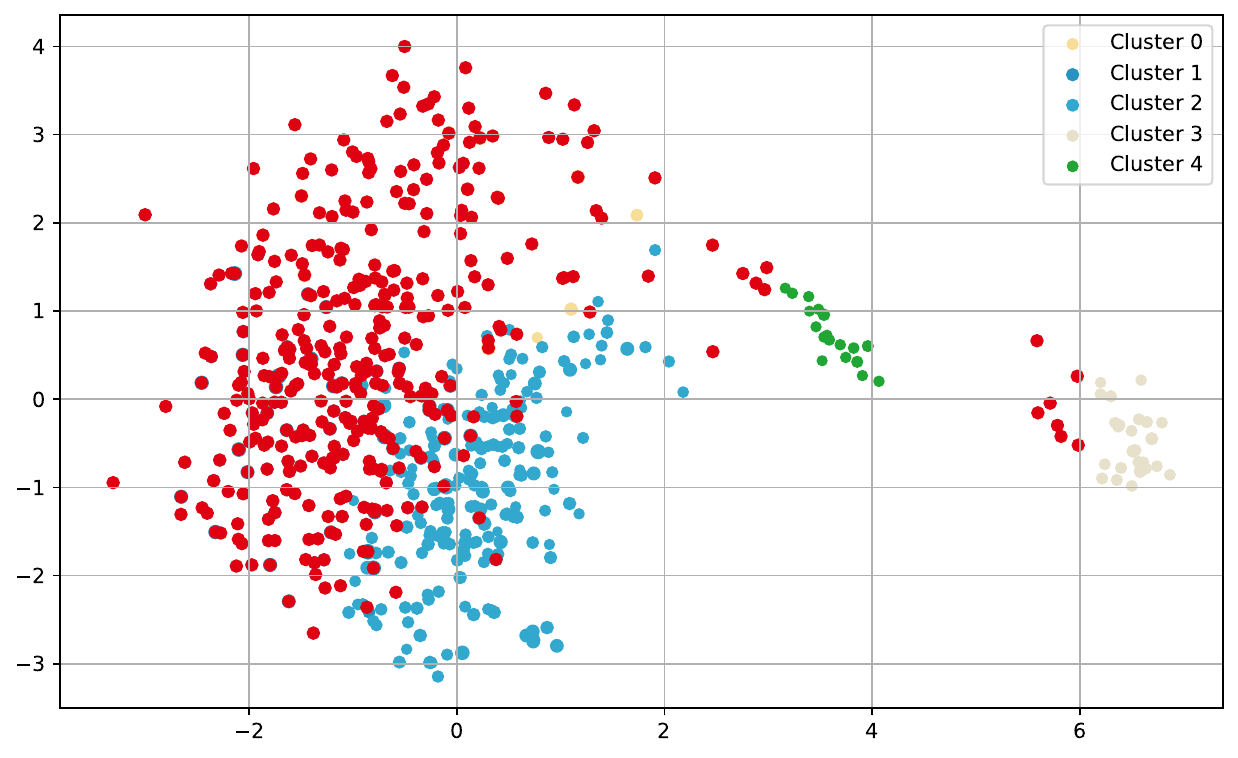}
        			\label{coprotector_ac_1}
        		\end{minipage}
        	}
	\newline % Start a new line for the third row of subfigures
         \subfigure[1st Round Results of CodeMark]{
        		\begin{minipage}[t]{0.47\linewidth} % 0.48 for two columns layout
        			\centering
        			\includegraphics[width=1\linewidth]{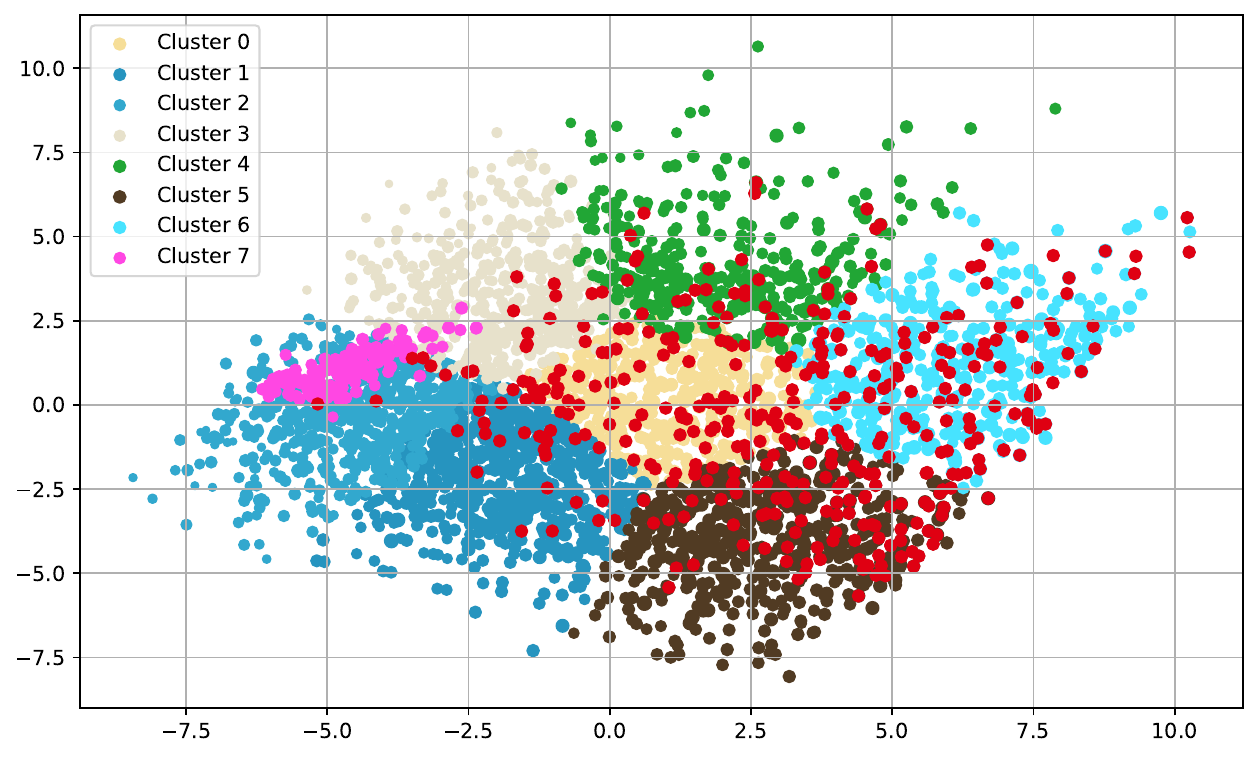}
        			\label{codemark_ac_0}
        		\end{minipage}
        	}
	\subfigure[2nd Round Results of CodeMark]{
		\begin{minipage}[t]{0.47\linewidth} % 0.48 for two columns layout
			\centering
			\includegraphics[width=1\linewidth]{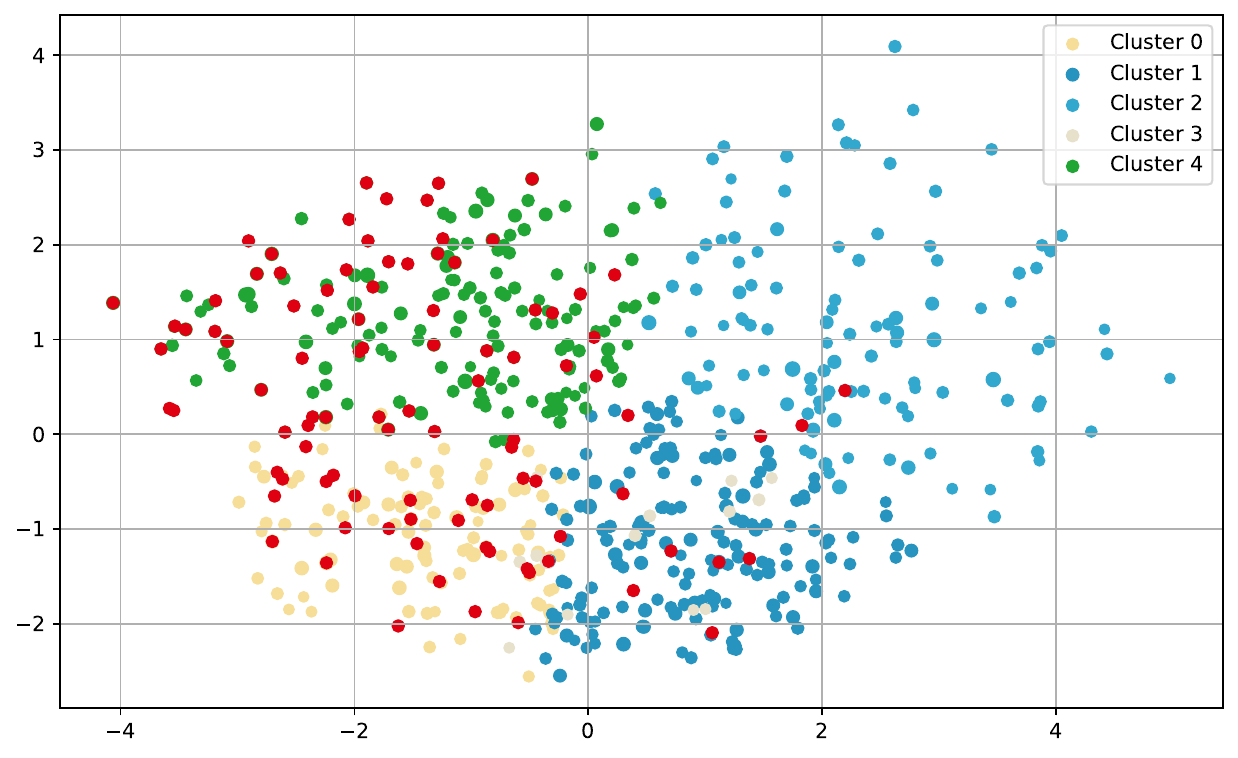}
			\label{codemark_ac_1}
		\end{minipage}
	}
 	\newline % Start a new line for the second row of subfigures
	\subfigure[1st Round Results of \sysname]{
		\begin{minipage}[t]{0.47\linewidth} % 0.48 for two columns layout
			\centering
			\includegraphics[width=1\linewidth]{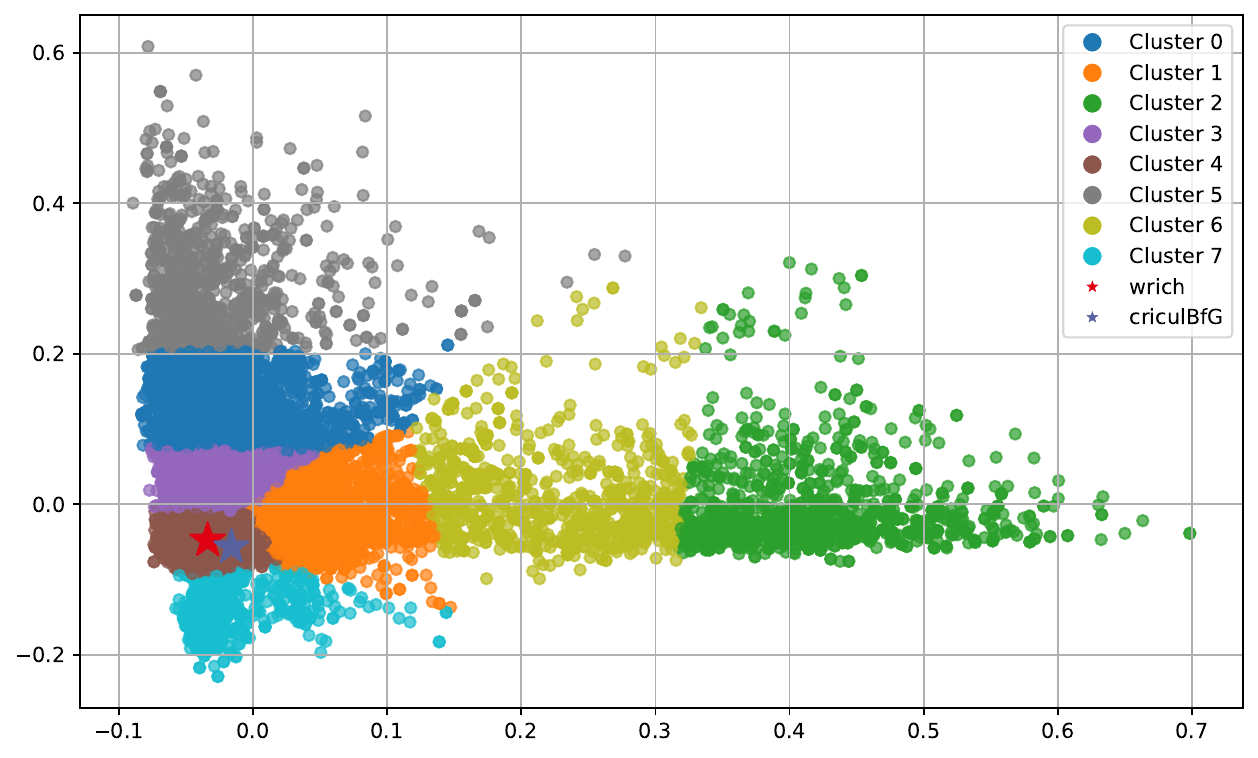}
			\label{ours_ac_0}
		\end{minipage}
	}
	\subfigure[2nd Round Results of \sysname]{
		\begin{minipage}[t]{0.47\linewidth} % 0.48 for two columns layout
			\centering
			\includegraphics[width=1\linewidth]{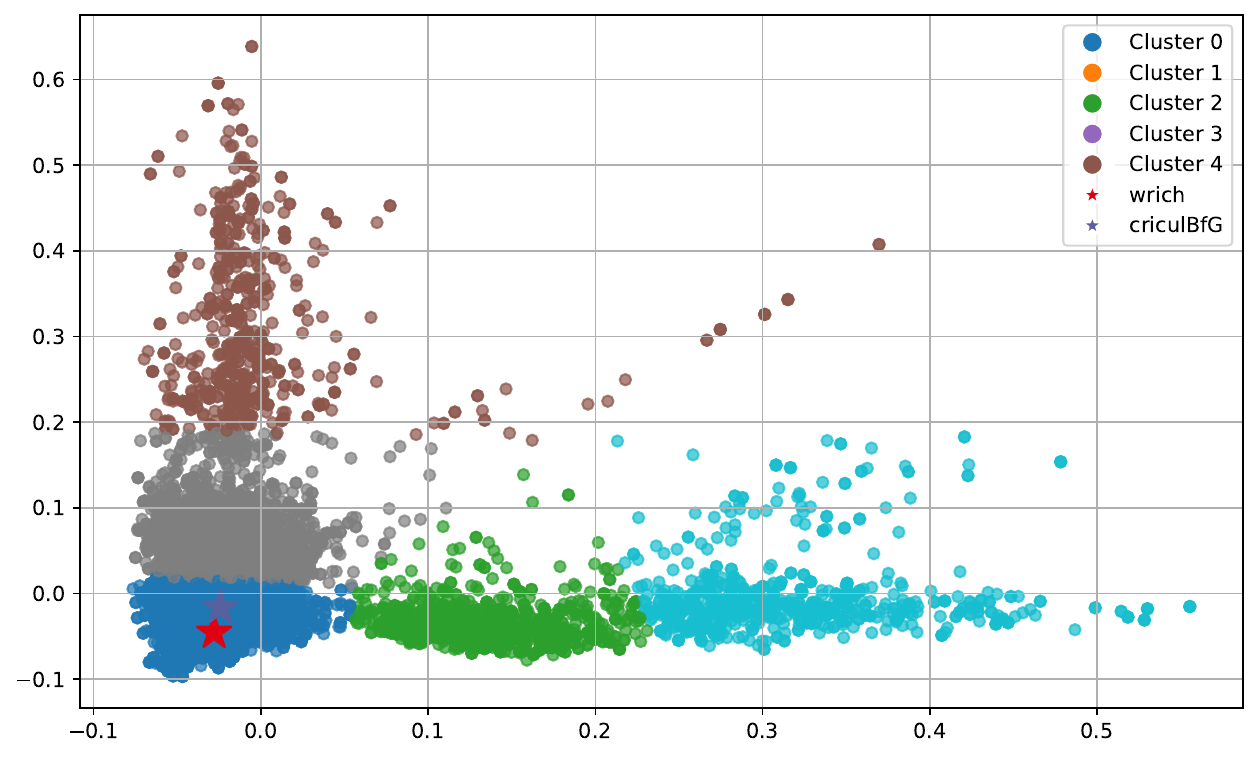}
			\label{ours_ac_1}
		\end{minipage}
	}
	\caption{Two-Round Clustering Results of CoProtector, CodeMark, and Our Method.}
 \vspace{-0.5cm}
	\label{AC}
\end{figure}

We employ the k-means clustering method to conduct clustering operations on CodeMark, CoProtector, and \sysname. However, as a dataset backdoor watermarking, our triggers are embedded within the model's tokenizer vocabulary. Therefore, we focus on clustering the vocabulary tokens rather than clustering the dataset itself. Due to limited computational resources, we are unable to cluster the entire dataset; instead, we randomly select 4,000 training samples from the original clean dataset. For CoProtector, we set a contamination rate of 20\%. For CodeMark, we choose the type with the highest watermark verification rate in each programming language as the trigger embedded in the samples. In our approach, we set up a tokenizer vocabulary embedded with Mark1 and Mark2 for clustering purposes. The clustering results are shown in Figure \ref{AC}, where we highlight the samples carrying triggers in red for the clustering results of CodeMark and CoProtector. In our clustering results, we mark the positions of the embedded watermark words with a star.

In our experiment, the first round of clustering was set to 8 categories. In the second round, the category with the highest number of trigger samples from the first round was selected and subdivided into 5 categories. The experimental results show that both CodeMark and CoProtector can successfully identify the categories containing trigger samples after two rounds of clustering. However, compared to CoProtector, CodeMark demonstrates greater robustness in its clustering approach, with the second round of clustering results showing that samples with triggers are mostly gathered in two classes. Compared to CoProtector, CodeMark requires more clustering rounds to locate the trigger samples. Due to the limited number of trigger samples and the introduction of noise, which results in the trigger words having similar vector representations to other words, our watermarked words are still categorized within the normal vocabulary classes after two rounds of clustering. This indicates that, compared to the baseline, under the same clustering setup, our method is more difficult to detect for trigger words, thus exhibiting better stealth.

\section{DataSet and Parameter details}
\label{Parameter}
\begin{table}[h]
  \caption{The volume of each programming language dataset.}
  \label{dataset}
\begin{tabular}{ccccccc}
\hline
      & Python & PHP    & Go     & Ruby  & JavaScript & Java   \\ \hline
Train & 251820 & 241241 & 167288 & 24927 & 58025      & 164923 \\
Valid & 13914  & 12982  & 7325   & 1261  & 3885       & 5183   \\
Test  & 14918  & 14014  & 8122   & 1400  & 3291       & 10955  \\ \hline
\end{tabular}
\end{table}
We conducted our experiments using the CodeXGLUE dataset, a large open-source dataset designed to support research in code search and related tasks. This dataset includes code examples from multiple programming languages such as Python, Java, JavaScript, PHP, Ruby, and Go, along with their corresponding natural language descriptions. The code examples encompass functions, classes, and other code snippets, covering a wide range of programming topics and application scenarios. The data volume for various languages in the dataset is listed in Table \ref{dataset}.

For computational resources, we utilized an NVIDIA GeForce RTX 4090 24GB graphics card. In terms of training configurations, we followed the initial setup of the CodeT5 model, employing the CodeT5-small pre-trained model. The maximum lengths of the source and target sequences were set to 512 and 256, respectively, with a linearly decaying peak learning rate of 2e-4. The batch size was set to 64, and training was halted if the model's performance did not improve after three consecutive rounds. No fixed number of training epochs was predefined.

\section{Algorithm}
The algorithm \ref{watermark_embedding_optimized} aims to embed a watermark into a tokenizer through the following process: First, the input code is converted into an Abstract Syntax Tree (AST), and all identifiers along with their positional information are extracted from the AST (lines 3-4). Next, an empty list is initialized to store code variants and their importance scores (line 5). Each identifier is then traversed, replaced with a placeholder UNK, and modified code is generated (lines 6-7). The modified AST is restored into a code string, and its importance score is calculated using a code summarization model, with the results stored in the variants list (lines 8-10). Subsequently, the code variant with the highest importance score is selected from the list (line 11), and the corresponding best identifier is extracted from the dataset (line 12). The best identifier is tokenized, and the lowest-frequency token is identified (lines 13-14). Noise is then added to this token (line 15). Finally, the merge rule associated with the noisy token is retrieved from the tokenizer, and the tokenizer is updated with the noisy token and merge rule to generate the watermarked tokenizer $T_m$, which is returned as the output (lines 16-18). 

\begin{algorithm}[ht]
\caption{Watermark Embedding Algorithm with Identifier Position Extraction}
\label{watermark_embedding_optimized}
\begin{algorithmic}[1] % 每一行都显示数字
\REQUIRE Code variant sets $C$, Dataset $D$, Code Summarization model and tokenizer $M, T$ % 输入参数
\ENSURE Watermarked Tokenizer $T_m$ % 输出结果
\STATE ast\_tree = convert\_to\_ast(code) % 转换代码为AST树
\STATE (identifiers, positions) = extract\_identifiers\_with\_positions(ast\_tree) % 提取标识符及其位置信息
\STATE variants = [] % 初始化变体列表
% 遍历标识符进行替换并计算重要性得分
\FOR{(identifier, position) \textbf{in} zip(identifiers, positions)}
    \STATE modified\_code = replace\_identifier\_at\_position(ast\_tree, identifier, position, 'UNK') % 替换为UNK
    \STATE variant\_code = restore\_code(modified\_code) % 还原代码
    \STATE importance\_score = model\_output(variant\_code) \COMMENT{Get model output score}
    \STATE variants.append((variant\_code, importance\_score))
\ENDFOR
\STATE best\_variant = max(variants, key=lambda x: x[1]) % 选择得分最高的变体
\STATE best\_identifier = extract\_identifier\_from\_dataset(D, best\_variant) % 提取最佳标识符
\STATE tokens = tokenizer.tokenize(best\_identifier) % 获取分词token
\STATE low\_freq\_token = find\_low\_frequency\_token(tokens) % 找到最低频token
\STATE noisy\_token = add\_noise(low\_freq\_token) % 加噪
\STATE merge\_rule = tokenizer.get\_merge\_rule(noisy\_token) % 获取合并规则
\STATE $T_m$ = update($T$, noisy\_token, merge\_rule) \COMMENT{Update word dictionary and merge file}
\STATE return $T_m$
\end{algorithmic}
\end{algorithm}

\end{document}